\documentclass[12pt]{article}
\begin{document}

\begin{center}
{\Large {\bf Quantization with maximally degenerate Poisson
brackets: The harmonic oscillator! } }\\[10mm]
{\Large {\bf Yavuz Nutku}}\\
Feza G{\"u}rsey Institute P. O. Box 6 \c{C}engelk{\"o}y, Istanbul
81220 Turkey \\
\end{center}

\noindent Nambu's construction of multi-linear brackets for
super-integrable systems can be thought of as degenerate Poisson
brackets with a maximal set of Casimirs in their kernel. By
introducing privileged coordinates in phase space these degenerate
Poisson brackets are brought to the form of Heisenberg's
equations. We propose a definition for constructing quantum
operators for classical functions which enables us to turn the
maximally degenerate Poisson brackets into operators. They pose a
set of eigenvalue problems for a new state vector. The requirement
of the single valuedness of this eigenfunction leads to
quantization. The example of the harmonic oscillator is used to
illustrate this general procedure for quantizing a class of
maximally super-integrable systems.

\vspace{1cm}



\vspace{1cm}

\section{Introduction}

The passage from classical to quantum mechanics is based on the
Hamiltonian formulation of the classical system. This is most apt
because one of the crowning achievements of Newtonian mechanics
was the construction of its underlying theory of symplectic
structure. The great works of Hamilton, Poisson, Jacobi and
Darboux in late $19^{th}$ century gave it the appearance of a
finished gem. However, it was only in the second half of $20^{th}$
century that we came to understand this theory to be much richer
and certainly very far from complete \cite{lich}. Nowhere is this
more manifest than in the theory of completely integrable systems
where we encounter more than one Poisson structure as a matter of
routine \cite{magri}. How are the modern developments in classical
theory of Poisson structure going to reflect on the problem of
quantization?

Nambu \cite{nambu} published an alternative Hamiltonian
formulation of super-integrable systems based on ideas he had as a
student, but only when he was already one of the leading
theoretical physicists of all time. The literature on the Nambu
bracket has followed Nambu's idea of regarding it as a
multi-linear object rather than bi-linear as in the case of
Poisson brackets. The problem of quantization with Nambu brackets
has been discussed in the framework of deformation quantization
\cite{lmpguys3} - \cite{czdirac}. In particular, we refer to the
recent work of Curtright and Zachos \cite{czbook} which serves as
a textbook for this line of research. There is also a geometrical
approach using complex projective Hilbert space endowed with
K\"ahler structure \cite{lane}-\cite{mike}. In this paper we shall
adopt a new approach, completely different from these.

We shall avoid multi-linear products altogether but instead focus
on taking Nambu brackets as Poisson brackets, albeit degenerate
ones because they admit a full set of Casimirs in their kernel.
The free Euler top that Nambu discussed as an illustration of his
ideas is a classic example of super-integrable systems that arise
from the existence of hidden symmetries. Super-integrability is
much stronger than Liouville integrability because such systems
are required to admit first integrals that number not half, but
only one less than the dimension of the dynamical system itself.
The discussion of the dynamics is then reduced to a single torus.
Each one of these conserved quantities can alternatively be taken
as {\it the} Hamiltonian function and each such choice results in
the construction of an independent degenerate Poisson bracket. In
this way we obtain the maximal number of compatible Poisson
structures for super-integrable systems. This new perspective
whereby the ideas of Nambu find realization as Poisson bracket was
discussed only recently \cite{gonu}, see also \cite{gunu} and
\cite{hoj}. To distinguish our approach from earlier literature it
would have been natural to use the name Nambu-Poisson bracket, but
this has been used before in another context and therefore we
shall simply call it maximally degenerate Poisson brackets.

Poisson tensors for super-integrable systems that are maximally
degenerate involve highly non-linear expressions of the dynamical
variables and we cannot naively carry them over to the quantum
mechanical domain. However, quantization with maximally degenerate
Poisson brackets becomes possible if we adopt the strategy of
introducing privileged coordinates in phase space and a new
definition for the quantum operator corresponding to any given
function of classical variables. In this way we construct quantum
operators for each maximally degenerate Poisson bracket and the
commutation relations are brought to the form of Heisenberg's
equations. Thus we set up a set of eigenvalue problems for a new
state vector. The quantization condition is the single valuedness
of this eigenfunction which is a new object, not immediately
related to Schr\"odinger's wave function.

We shall not present a formalism but instead give an outline of
our approach interspersed with an illustrative example which is
the simplest super-integrable system, namely the harmonic
oscillator in two dimensions! This will help to fix ideas. Even
for this simplest system the calculations are rather involved.
Later on we shall consider the general formalism and other
familiar problems which admit maximally degenerate Poisson
structure, such as the Kepler problem \cite{chat}, the example of
$S^2$ that Curtright and Zachos \cite{cz} have discussed and
Calogero-type systems \cite{gonu} as well as the completely
integrable Smorodinsky-Winternitz potentials \cite{sw}, \cite{stw}
of the Schr\"odinger equation \cite{ns}.

\section{Classical maximally degenerate Poisson \\ brackets}

We start with a $2 n$-dimensional Hamiltonian system with the
equations of motion given by
\begin{equation}
\dot{X}^A = [X^A, H^1]_0 = J^{AB}_0 \frac{\partial H^1}{\partial
X^B} \label{hameqs}
\end{equation}
where $ X^A \in \{q^1, ..., q^n,p_1, ...,p_n\} $ and we shall
consider all the canonical variables on the same footing. Capital
Latin indices will range over $2 n$ values. In equations
(\ref{hameqs}) we have the canonical Poisson tensor
\begin{equation}
J_0 =\left(\begin{array}{cc} 0&I \\-I&0
\end{array} \right) \label{canonical}
\end{equation}
in Darboux form and $[,]_0$ is the usual Poisson bracket. Assuming
these equations of motion to be super-integrable, we shall have $2
n - 2$ further first integrals of motion in addition to the usual
Hamiltonian function $H^1$. Every maximally super-integrable
system admits a set of first integrals $ H^\alpha$ where Greek
indices will be reserved to range over $2n-1$ values.

Due to the functional independence of these first integrals, their
gradients $\partial H^{\alpha} / \partial X^A$ define $2 n - 1$
linearly independent vectors orthogonal to the velocity vector. By
taking their full cross-product we determine at each point in
phase space a unique direction which is precisely that of the
velocity vector. Then the trajectory is determined by the
equations of motion in the Nambu form
\begin{eqnarray}
\dot{X}^{A} &=& \frac{\partial(H^1,...  ..., H^{2n-1})}{ \partial(
X^1,...\widetilde{X^A}... X^{2n})} \nonumber \\
&=& \varepsilon_{\alpha_1...\alpha_{2n-1}}
\epsilon^{A_1...(A)...A_{2n}} \; \frac{\partial H^{\alpha_1}
}{\partial X^{A_1} } \; ...\; \frac{\partial
H^{\alpha_{2n-1}}}{\partial X^{A_{2n} } } \label{nambueqs}
\end{eqnarray}
where tilde over a quantity indicates that it will be omitted and
indices enclosed by round parentheses are excluded from the
implied summation. Here $\epsilon^{A_1...A_{2n}}$ with
\begin{equation}
\epsilon^{1 2 3 ... 2n} = \frac{1}{\sqrt{g}} \label{eps}
\end{equation}
is the $2n$-dimensional completely anti-symmetric Levi-Civita
tensor density in phase space, whereas
$\varepsilon_{\alpha_1...\alpha_{2n-1}}$ is the permutation symbol
in $2n-1$ dimensions. The factor of proportionality $\sqrt{g}$
will be determined from the requirement that both the magnitude as
well as the direction of the velocity vector in phase space is
given by the original equations of motion. It will be recognized
as the volume density in phase space. For the class of
super-integrable systems we shall consider here $\sqrt{g}$ will be
time-independent, that is, a function of integrals of motion
$H^\alpha$ only. See \cite{gunu} for cases where this fails and
Nambu mechanics is generalized.

In Nambu's form of the equations of motion (\ref{nambueqs}) it is
immediately evident that they can be expressed in bracket form in
$2n-1$ different ways
\begin{equation}
\dot{X}^A = \left[X^A, H^\alpha \right]_\alpha =  J^{AB}_\alpha
\frac{\partial H^\alpha}{\partial X^B} \label{hameqss}
\end{equation}
depending on which one of the conserved quantities $H^\alpha$ we
take as the Hamiltonian function. In (\ref{hameqss}) there is no
summation implied over $\alpha$ which simply enumerates the
different ways the original equations of motion can be written in
maximally degenerate Poisson bracket form. The labels of both the
Hamiltonian function and the degenerate Poisson bracket must
coincide. The expression for the $2n-1$ maximally degenerate
Poisson tensors will then be
\begin{equation}
 J^{AB}_\alpha =  \varepsilon_{\alpha_1...(\alpha) ...\alpha_{2n-1} }
 \epsilon^{A_1... (A)... (B)...A_{2n} } \;
 \frac{\partial H^{\alpha_1}}{\partial X^{A_1} }...
 \widetilde{ \frac{\partial H^\alpha}{\partial X^{B} }    }...
\frac{\partial H^{\alpha_{2n-1} }}{\partial X^{A_{2n}} }
\label{poissons}
\end{equation}
whereby (\ref{hameqss}) and (\ref{poissons}) together will simply
yield (\ref{nambueqs}). This point follows from \cite{gonu}, see
also \cite{gunu} and \cite{hoj}. Note that the equations of motion
can be obtained in two different ways using the same Hamiltonian
function $H^1$ either with Darboux's canonical Poisson tensor
(\ref{canonical}), or the first one of the degenerate Poisson
tensors constructed according to (\ref{poissons}) with $\alpha=1$.
The proof of the Jacobi identity
\begin{equation}
 J^{M[A}_\alpha  \frac{\partial J^{BC]}_{\beta}}{\partial X^M}
  =0 \label{jacobi}
\end{equation}
where square brackets denote complete skew-symmetrization and the
fact that all these Poisson tensors are degenerate
\begin{equation}
\det J^{AB}_\alpha = 0 \label{degenerate}
\end{equation}
can be found in \cite{gonu}. From the construction
(\ref{poissons}) it is manifest each $J_\alpha$ admits $2n-2$
Casimirs $H^\beta$ with $\alpha \ne \beta$. Thus in all cases the
rank of the Poisson tensors $J^{AB}_\alpha$ will be $2$
independent of the dimension of the dynamical system. These
maximally degenerate Poisson tensors are compatible and thus form
a Poisson pencil, a fact manifest from the lack of any restriction
we have put on $\alpha$ and $\beta$ in the Jacobi identity
(\ref{jacobi}). However, none of the degenerate Poisson tensors is
compatible with the Poisson tensor (\ref{canonical}) in Darboux's
canonical form. The ``basic identity" discussed in the literature
on Nambu brackets is simply the Jacobi identity (\ref{jacobi}) for
maximally degenerate Poisson brackets.

We shall now present the maximally degenerate Poisson tensors for
the harmonic oscillator in two dimensions, $n=2$, as the simplest
non-trivial example of this subject. The equations of motion are
given by the usual Hamiltonian function
\begin{equation}
H^1 =  \frac{1}{2 m} \left( p_x^2 + p_y^2 \right) + \frac{1}{2} k
(x^2+y^2) \label{hamgen}
\end{equation}
in the form of Hamilton's equations with Darboux's Poisson tensor
(\ref{canonical}). But the harmonic oscillator is
super-integrable, that is, it admits two further first integrals
\begin{eqnarray}
H^2 & = & \frac{1}{2 m} ( p_x^2 - p_y^2 ) +\frac{1}{2} k ( x^2 -
y^2 )  \label{h12}\\
H^3 & = & \frac{1}{2}  L^2, \qquad L  \equiv  x  p_y - y  p_x
\label{h13}
\end{eqnarray}
where $H^3$ is the squared magnitude of angular momentum and $H^2$
is the super-integral which comes from the fact that the equations
of motion decouple in Cartesian coordinates. This choice of first
integrals is such that we can generalize our discussion \cite{ns}
to completely integrable Smorodinsky-Winternitz potentials
\cite{stw} of the Scr\"odinger equation in two dimensions. The
volume density is given by
\begin{equation}
\sqrt{g} = 2 \, L \, M, \qquad M \equiv  \frac{1}{m} p_x p_y + k x
y =\sqrt{(H^1)^2 - (H^2)^2 - 2 \omega^2 H^3}
 \label{vol}
\end{equation}
which is manifestly a conserved quantity. Here $\omega^2=k/m$ is
the frequency of oscillation. Perhaps in (\ref{vol}) we should
have used the symbol $(H^2)'$ instead of $M$ because if we had
started with $M$ as the super integral, then our $H^2$ in
(\ref{h12}) would enter into the volume factor. The three
degenerate Poisson tensors that follow from the construction
(\ref{poissons}) are given by
\begin{eqnarray}
\left[x,y\right]_1  =  \frac{1}{ 2 m M} \left( x p_x - y p_y
\right), &\qquad&  \left[p_x, p_y\right]_1  =  - \frac{k}{ 2 M }
\left(
x p_x - y p_y \right),  \nonumber \\
\left[x, p_x \right]_1  =  \frac{1}{2}, &\qquad& \left[y, p_y
\right]_1
=  \frac{1}{2},  \label{nbpmosc1}\\
 \left[x, p_y \right]_1 = \frac{1}{2 M}  \left( \frac{1}{m} p_x^2
 + k y^2 \right), &\qquad& \left[y, p_x \right]_1  =   \frac{1}{2 M}
 \left( \frac{1}{m} p_y^2 + k x^2 \right) , \nonumber
\end{eqnarray}
together with
\begin{eqnarray}
\left[x,y\right]_2 =  - \frac{1}{ 2 m M} \left( x p_x + y p_y
\right), &\qquad&  \left[p_x, p_y\right]_2  =   \frac{k}{ 2 M }
\left(
x p_x + y p_y \right), \nonumber\\
\left[x, p_x \right]_2  = \frac{1}{2}, &\qquad& \left[y, p_y
\right]_2
=  - \frac{1}{2}, \label{nbpmosc2}\\
\left[x, p_y \right]_2  =  - \frac{1}{2 M}  \left( \frac{1}{m}
p_x^2 - k y^2 \right), &\qquad& \left[y, p_x \right]_2  =
\frac{1}{2 M} \left( \frac{1}{m} p_y^2 - k x^2 \right) \nonumber
\end{eqnarray}
and the third one is the simplest
\begin{eqnarray}
\left[x,y\right]_3 = - \frac{1}{m^2 L M }\; p_x p_y, & \qquad &
\left[p_x, p_y\right]_3 = - \frac{k^2}{ L M }\; x y, \nonumber
\\[2mm]
\left[x, p_x \right]_3 = 0, & \qquad & \left[y, p_y \right]_3 = 0,
 \label{nb3osc}  \\
\left[x, p_y \right]_3 =  \frac{\omega^2}{ L M }\;  y p_x , &
\qquad & \left[y, p_x \right]_3 = - \frac{\omega^2}{L M}
 \; x p_y \nonumber
\end{eqnarray}
and has an interesting free particle limit $\left[x,y\right]_3 = -
1/ m L$ with all others vanishing. This is reminiscent of the
Dirac bracket \cite{dirac} in the Landau problem for the motion of
a charged particle in a strong magnetic field with angular
momentum replacing the magnetic field. Curtright and Zachos
\cite{czdirac} have discussed the relationship between Dirac and
Nambu brackets.

We see that the maximally degenerate Poisson tensors constructed
according to (\ref{poissons}) are highly nonlinear which will pose
nasty problems if we were to carry them over to the quantum
mechanical domain naively. However, there is a way to overcome
this difficulty. By introducing new coordinates in phase space
that consist of $\{ H^\alpha \}$ and $H^{2n}$ which is a ``time"
variable
\begin{equation}
\frac{d H^{2n}}{d t} = 1 \label{h44}
\end{equation}
the Nambu-Poisson bracket relations can be summed up in the form
\begin{eqnarray}
\left[H^\alpha, H^\beta\right]_\gamma  & = & 0 \qquad \forall \;\;
\alpha, \beta, \gamma \label{sum1}  \\
\left[ H^{2n}, H^\alpha \right]_\beta & = & \delta^\alpha_\beta
\label{sum2}
\end{eqnarray}
of Heisenberg's equations.

For the harmonic oscillator the choice
\begin{equation}
H^4 = \frac{1}{\omega} \tan^{-1}  \left( m \omega \frac{ x + y }{
p_x + p_y } \right) \label{h4}
\end{equation}
satisfies (\ref{h44}). We note that this choice of $H^4$ is not
unique, a point which will emerge as immaterial in what follows
because we shall only need its gradients and they are directly
obtained from the equations of motion. It requires the usual
straight-forward but lengthy calculations to verify the properties
(\ref{sum1}), (\ref{sum2}) in the case of the harmonic oscillator.

Another advantage of introducing privileged coordinates in phase
space is that now a Riemannian metric on phase space is suggested.
In the example of the harmonic oscillator this is given by
\begin{eqnarray}
d s^2 & = &  2 H^1 d H^1 d H^3 + 2 H^2 d H^2 d H^3 + 2 \omega^2 d
(H^3)^2 \label{wow} \\
& & + H^3 \left[ d (H^1)^2 - d (H^2)^2 \right] + d (H^4)^2
\nonumber
\end{eqnarray}
which has the determinant given by (\ref{vol}).

\section{Quantum operators for maximally degenerate Poisson brackets}

In elementary quantum mechanics courses we are taught that the
passage to quantum mechanics requires
\begin{equation}
[p_i, q^k]_0 = \delta^k_i \qquad \longrightarrow \qquad \hat p_i
\hat q^k - \hat q^k \hat p_i = \frac{h}{i} \delta_i^k \label{prop}
\end{equation}
where hats denote operators, and shown Schr\"odinger's solution to
it
\begin{equation}
\hat p_i = \frac{\hbar}{i} \frac{\partial}{\partial q^i}, \qquad
\hat q^k = q^k \label{schsln}
\end{equation}
which always mystifies. Now we shall propose a definition which
makes sense out of this.

{\bf Definition}: {\it Given a Poisson tensor} $J$ {\it and} {\bf
any} {\it function of the canonical variables}  $F$, {\it then the
quantum operator} $\hat F$ {\it that corresponds} to $F$ {\it is
given by}
\begin{equation}
\hat F =  \frac{\hbar}{i} \, J^{A B} \frac{\partial F}{\partial
X^A} \frac{\partial}{\partial X^B} \label{definition}
\end{equation}
{\it which is proportional to the Poisson vector-field appropriate
to} $F$.

With this definition we can immediately see that Schr\"odinger's
solution (\ref{schsln}) is obtained from Darboux's Poisson tensor
(\ref{canonical}) and the functions $p_i$. Of course we do not
apply this definition to the functions $q^i$ but instead take them
to be $c$-numbers, because then $\hat p_i$ and $\hat q^i$ would
commute contradicting (\ref{prop}).

This is not the appropriate place to discuss the merits of the
definition (\ref{definition}), or any lack thereof. We shall
instead apply it to the degenerate Poisson brackets and see if we
arrive at a consistent theory.

After staring at (\ref{sum1}) and (\ref{sum2}) for a while, it
becomes clear that in the case of quantization with degenerate
Poisson brackets we need to regard each $H^\alpha$ as a $c$-number
and construct the quantum operators using $H^{2 n}$. There will be
$2 n - 1$ such operators
\begin{equation}
\hat H^{2n}_\alpha =  \frac{\hbar}{i} \, J^{A B}_\alpha
\frac{\partial H^{2n}}{\partial x^A} \frac{\partial}{\partial x^B}
\label{h4ops}
\end{equation}
corresponding to the full set of degenerate Poisson brackets. All
of these operators must commute
\begin{equation}
[\hat H^{2n}_\alpha , \hat H^{2n}_\beta ] = 0 \qquad \forall \;
\alpha , \beta \label{h4scommute}
\end{equation}
or more precisely, their Lie brackets must vanish. These are very
strong requirements, but they will be satisfied because we are
dealing with maximally super-integrable systems.

Let us illustrate this with the example of the quantum operators
for the harmonic oscillator. From the definition (\ref{h4ops})
using the degenerate Poisson brackets (\ref{nbpmosc1})-(\ref
{nb3osc}) with $H^4$ given by (\ref{h4}), we arrive at the
following quantum operators
\begin{equation}
{\hat H}^4_1 = \frac{\hbar}{2 i L M} \left( y
\frac{\partial}{\partial x} + x \frac{\partial}{\partial y} + p_y
\frac{\partial}{\partial p_x} + p_x \frac{\partial}{\partial p_y}
\right), \label{op1}
\end{equation}
\begin{eqnarray}
\hat H^4_2 & = & \frac{\hbar}{4 i M  ( H^1 +M )} \left\{
 - \frac{1}{m} (p_x^2-p_y^2) \left( y \frac{\partial}{\partial
x} + x \frac{\partial}{\partial y} \right)  \right. \nonumber \\
&& +  k (x+y)^2  \left( y \frac{\partial}{\partial x} - x
\frac{\partial}{\partial y} \right) - k (x^2-y^2) \left( p_y
\frac{\partial}{\partial p_x} + p_x \frac{\partial}{\partial p_y}
\right) \label{op2} \\
&&  +\frac{2}{m} p_x p_y (x+y) \left( \frac{\partial}{\partial x}
-\frac{\partial}{\partial y} \right)
 + 2 k x y (p_x+p_y) \left( \frac{\partial}{\partial
p_x}- \frac{\partial}{\partial p_y} \right) \nonumber\\
  && \left.   + \frac{1}{m} (p_x+p_y)^2 \left( p_y
\frac{\partial}{\partial p_x} - p_x \frac{\partial}{\partial p_y}
\right)  \right\} \nonumber
\end{eqnarray}
and
\begin{eqnarray}
{\hat H}^4_3 & = &  \frac{\hbar}{2 i L M (H^1+M) } \left\{
\frac{}{} \right. \nonumber
\\ &&   \left[ k y ( x+y) + \frac{1}{m} p_y ( p_x+p_y) \right]
\left( \frac{1}{m} p_x \frac{\partial}{\partial x} - k x
\frac{\partial}{\partial p_x} \right)  \label{op3} \\
&& \left. - \left[ k x ( x+y) + \frac{1}{m} p_x ( p_x+p_y)
\right]\left( \frac{1}{m} p_y \frac{\partial}{\partial y} - k y
\frac{\partial}{\partial p_y} \right) \right\}. \nonumber
\end{eqnarray}
It is a straight-forward but this time most lengthy calculation to
verify that all Lie brackets of (\ref{op1}), (\ref{op2}) and
(\ref{op3}) indeed do vanish.

We note that earlier Hietarinta \cite{jarmo} had discussed Nambu
tensors and commuting vector fields. Hietarinta's vector fields
are just the Hamiltonian vector fields obtained from Darboux's
Poisson tensor (\ref{canonical}). It is interesting to compare the
operators (\ref{op1}) - (\ref{op3}) to Hietarinta's Hamiltonian
vector fields for the harmonic oscillator
\begin{eqnarray}
 \nonumber \\[2mm]
H^1 &=& - \frac{1}{m} p_x \frac{\partial}{\partial x} + k x
\frac{\partial}{\partial p_x} - \frac{1}{m} p_y
\frac{\partial}{\partial y} + k y \frac{\partial}{\partial p_y}
\label{j1} \\
 H^2 &=& - \frac{1}{m} p_x \frac{\partial}{\partial
x} + k x \frac{\partial}{\partial p_x} + \frac{1}{m} p_y
\frac{\partial}{\partial y} - k y \frac{\partial}{\partial p_y}
\label{jarm} \\
H^3 &=& L \left( y \frac{\partial}{\partial x} - x
\frac{\partial}{\partial y} + p_y \frac{\partial}{\partial p_x} -
p_x \frac{\partial}{\partial p_y} \right) \label{j3}
\end{eqnarray}
which consist of the generators of three commuting rotations in
four dimensions. Finally, we note that the operators $ \hat
H^\alpha_\alpha $ obtained from Nambu-Poisson brackets in
accordance with the definition (\ref{definition}) reduce to $\hat
H^1_1 = \hat H^2_2 = \hat H^3_3 = H^1$ of (\ref{j1}).

We have seen that the maximally degenerate Poisson quantum
operators assume formidable expressions even in the case of the
simplest possible classical system. To quantize we need to set up
$2n-1$ eigenvalue problems
\begin{equation}
\hat H^{2n}_\alpha \Phi  = \lambda_\alpha \Phi \label{opeqs}
\end{equation}
for some state vector $\Phi$ which is {\bf not} the Schr\"odinger
wave function. This looks like a very difficult problem, but its
solution is simplicity itself
\begin{equation}
\Phi  =  \exp {\frac{i}{\hbar} \Sigma_{\alpha=1}^{2n-1}
\lambda_\alpha H^\alpha } \label{psi}
\end{equation}
where we recall that $H^\alpha$ are the privileged coordinates in
phase space\footnote{This result has a counterpart in usual
quantum mechanics. For the Schr\"odinger wave function, the
solution of the eigenvalue problem $$\hat p_i \Psi = p_i \Psi $$
is the plane wave $$\Psi  = \exp {\frac{i}{\hbar} \Sigma_{i=1}^n
p_i q^i } $$ quite analogous to (\ref{opeqs}) and (\ref{psi}).
This parallel shows the virtue of our definition
(\ref{definition}) for turning classical functions into quantum
operators. There is, however, no resemblance between the state
vectors $\Phi$ and $\Psi$ in terms of their physical meaning.}.
But these are also conserved quantities and therefore
\begin{equation}
\dot \Phi = 0
\end{equation}
and we have a frozen time formalism. In particular, from
(\ref{opeqs}) and the expression for the eigenfunction (\ref{psi})
we find that for the case of the harmonic oscillator $\lambda_1$
is the energy, $\lambda_3$ is the magnitude of the angular
momentum and $\lambda_2$ is the constant of separation in the
Schr\"odinger equation.

The quantization condition is simply the requirement of single
valuedness of this eigenfunction. That is, $\Phi$ must be periodic
in the privileged coordinates $H^\alpha$. This forces the
eigenvalues $\lambda_\alpha$ to become integers. The specific
nature of these eigenvalues will be determined by the holonomy
structure of $\Phi$ which must be done on a case by case basis.

\section{Conclusion}

Developments in the theory of Poisson structure have not yet made
their full impact on quantum mechanics as they certainly will. In
this paper we have considered only one but an important aspect of
these developments, namely the maximal set of degenerate Poisson
brackets derived from Nambu's form of the equations of motion for
super-integrable systems. We have shown that these Poisson
brackets define skew-symmetric tensors of second rank satisfying
the Jacobi identity. The highly non-linear expressions we find for
these brackets become Heisenberg equations when we introduce new
coordinates in phase space that consist of the first integrals of
motion and a ``time" variable. We have proposed a definition for
turning classical functions into quantum operators and with its
help formulated a set of eigenvalue problems for super-integrable
systems. We emphasize again that the eigenfunction we have
introduced is an entirely new object that has nothing to do with
Schr\"odinger's wave-function. The solution for the eigenfunction
results in a phase factor which consists of a linear superposition
of the privileged coordinates on phase space. The requirement of
its single-valuedness is the quantization condition. The
determination of the precise nature of the eigenvalues is based on
the holonomy structure of $\Phi$.

This is a general procedure for alternative quantization of
maximally super-integrable systems. However, it is not the full
story because even in dynamical systems with three degrees of
freedom that Nambu first discussed, there are cases where the
factor of proportionality $\sqrt g$ is not conserved and Nambu
mechanics must be generalized \cite{gunu}.

\section{Acknowledgements}

As a student I was privileged to attend some courses taught by
Professor Nambu. He always emphasized main ideas and did so in
strikingly original ways. I hope this will serve as a token of my
appreciation. I would also like to thank Professor Segre for
teaching me a beautiful first real quantum mechanics course, in
spite of the fact that I had been struggling to understand it ever
since.

I thank the referees of this paper for interesting critical
comments. This work was in part supported by T\"UBA, Turkish
Academy of Sciences.

\end{document}